\documentclass[aps,chaos,twocolumn]{revtex4}   

\usepackage{graphicx,bm,hyperref,amssymb,amsmath}

\newlength{\wid}                       
\newcommand{\bi}{\begin{itemize}}
\newcommand{\ei}{\end{itemize}}
\newcommand{\ben}{\begin{enumerate}}
\newcommand{\een}{\end{enumerate}}
\newcommand{\be}{\begin{equation}}
\newcommand{\ee}{\end{equation}}
\newcommand{\bea}{\begin{eqnarray}} 
\newcommand{\eea}{\end{eqnarray}}
\newcommand{\bc}{\begin{center}}
\newcommand{\ec}{\end{center}}
\newcommand{\ie}{{\it i.e.\ }}
\newcommand{\eg}{{\it e.g.\ }}

\newcommand{\tbox}[1]{{\mbox{\tiny #1}}}
\newcommand{\mbf}[1]{{\mathbf #1}}
\newcommand{\etal}{{\it et.\ al.\ }}
\newcommand{\dr}{{d\mbf{r}}}    
\newcommand{\ofr}{(\mbf{r})}
\DeclareMathOperator{\vol}{vol}
\newcommand{\nd}[1]{{\partial_n\phi_{#1}}}      
\newcommand{\pO}{{\partial\Omega}}
\newcommand{\rO}{{\Omega'}}                     
\newcommand{\prO}{{\partial\Omega'}}            
\newcommand{\poin}{Poincar\'{e}\ }
\newcommand{\cre}{\ensuremath{a^{\dagger}}}      
\newcommand{\eps}{\varepsilon}

\begin{document}

\title{Quantum mushroom billiards}
\author{Alex H. Barnett}
\affiliation{Department of Mathematics, Dartmouth College, Hanover,
NH, 03755, USA}
\email{ahb(AT)math.dartmouth.edu}
\author{Timo Betcke}
\affiliation{School of Mathematics, The University of Manchester,
  Manchester, M13 9PL, UK}
\email{timo.betcke(AT)manchester.ac.uk}
\date{\today}

\begin{abstract}
We report the first large-scale statistical study of
very high-lying eigenmodes (quantum states) of the
mushroom billiard
proposed by L. Bunimovich in this journal, {\bf 11}, 802 (2001).
The phase space of this mixed system is unusual in that it has
a single regular region and a single chaotic region, and no KAM hierarchy.
We verify Percival's conjecture to high accuracy ($1.7\%$).
We propose a 
model for dynamical tunneling and
show that it predicts well the 
chaotic components of predominantly-regular
modes.
Our model explains our observed
density of such superpositions dying as $E^{-1/3}$ ($E$ is the eigenvalue).
%
We compare eigenvalue spacing distributions against
Random Matrix Theory expectations,
using $16000$ odd modes (an order of magnitude more than
any existing study).
We outline new variants of
mesh-free boundary collocation methods which enable us
to achieve high accuracy
and such high mode numbers
orders of magnitude faster than with competing methods.
\end{abstract}

\maketitle





{\bf
Quantum chaos is the study of the quantum (wave)
properties of Hamiltonian systems whose classical (ray) dynamics
is chaotic.
Billiards are some of the simplest and most studied examples;
physically their wave analogs are
vibrating membranes,
quantum,
electromagnetic, or acoustic cavities.
They continue to provide a wealth of theoretical challenges.
In particular `mixed' systems, where ray
phase space has both regular and
chaotic regions (the generic case),
are difficult to analyse.
Six years ago Bunimovich described~\cite{mushroom} a
mushroom billiard with simple mixed dynamics
free of the usual island hierarchies
of Kolmogorov-Arnold-Moser (KAM).
He
concluded by anticipating the
{\em growth of ``quantum mushrooms''}: it is this
gardening task that we achieve here, by developing
advanced numerical methods to collect an unprecedented
large number $n$ of eigenmodes
(much higher than competing numerics~\cite{demenez07} or
microwave studies~\cite{dietz}).
Since uncertainties scale as $n^{-1/2}$, a large $n$ is
vital for accurate spectral statistics and for studying the
semiclassical (high eigenvalue) limit.
We address three main issues. i) The conjecture of Percival~\cite{percival}
that semiclassically modes
live exclusively in invariant (regular or chaotic)
regions,
and occur in proportion to the phase space volumes.
ii) The mechanism for dynamical tunneling,
or quantum coupling between classically-isolated phase space regions.
iii) The distribution of spacings of nearest-neighbor eigenvalues,
about which recent questions have been raised~\cite{dietz}.
We show many pictures of modes,
including the boundary phase space (the so-called Husimi
function).
%
}

\begin{figure}
\bc
\includegraphics[width=\wid]{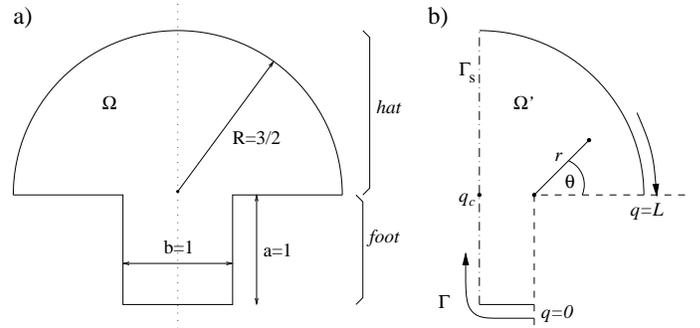}
\ec
\caption{
a) Mushroom billiard $\Omega$ used in this work.
The dotted line shows the reflection symmetry.
b) Desymmetrized half-mushroom $\rO$
used for mode calculation, and polar coordinates.
Dashed lines meeting at this corner
are zeros enforced by basis functions.
The remaining part of $\prO$ is $\Gamma$, comprising two pieces:
Dirichlet boundary conditions on the parts shown as solid,
while boundary conditions vary (see text) on the dash-dotted vertical line
$\Gamma_s$.
Boundary coordinate $q\in[0,L]$ parametrizes $\Gamma$.
\label{fig:geom}
}
\end{figure}

\section{Introduction}
\label{sec:i}

The 
nature of eigenfunctions of linear partial differential operators
in the short wavelength, or semiclassical, limit remains a
key open problem which continues to engage mathematicians and physicists
alike.
When the operator is the quantization of a classical Hamiltonian dynamical
system, the behavior of eigenfunctions
depends on the class of dynamics. In particular, hyperbolic dynamics
(exponential sensitivity to initial conditions, or chaos)
leads to irregular eigenfunctions, the study of which forms
the heart of a field known as `quantum chaos' \cite{gutz} or
`quantum ergodicity' \cite{debievre,zencyc}.
The planar billiard, or particle undergoing elastic reflection
in a cavity $\Omega\subset \mathbb{R}^2$, is one of the simplest examples.
Billiards exhibit a menagerie of dynamical
classes~\cite{disperse,stadium,whatbil}
ranging from
complete integrability (ellipses and rectangles)
to complete ergodicity (\eg Sinai billiard).
Bunimovich introduced the `mushroom' billiard~\cite{mushroom, porter}
with the novelty of a well-understood divided
phase-space comprising a single
integrable (KAM) region and a single ergodic region
\footnote{We note that a related Penrose-Lifshits mushroom construction
\cite{rauch} continues to find use in isospectral problems~\cite{fulling}}.
As seen in Fig.~\ref{fig:geom}a,
the mushroom is the union of a half-disk (the `hat') and a rectangle (the
`foot'); only trajectories reaching the foot are chaotic.
The simplicity of its phase space has allowed analysis of
phenomena such as
`stickiness' (power-law decay of correlations) in the ergodic
region~\cite{sticky1,sticky2}.

The quantum-mechanical analog of billiards
is the spectral problem of the Laplacian in $\Omega$
with homogeneous boundary conditions (BCs).
Choosing Dirichlet BCs (and units such that $\hbar=2m=1$) we have
\bea
        -\Delta \phi_j &=& E_j \phi_j \qquad \mbox{in}\; \Omega,
\label{eq:helm}
        \\
        \phi_j&=&0 \qquad\quad\mbox{on}\; \pO.
\label{eq:bcs}
\eea
This `drum problem' has a wealth of applications
throughout physics and engineering~\cite{KS}.
Eigenfunctions (or eigenmodes, modes)
$\phi_j$
may be chosen real-valued and orthonormalized,
$        \langle \phi_i, \phi_j\rangle \; := \;
        \int_\Omega \phi_i\ofr \phi_j\ofr \dr
        \;=\;\delta_{ij}$,
where $\dr:=dxdy$ is the usual area element.
`Energy' eigenvalues $E_1<E_2\le E_3\le \cdots \to\infty$
may be written $E_j = k_j^2$, where the (eigen)wavenumber $k_j$
is $2\pi$ divided by the wavelength.

Traditional numerical
methods to compute eigenvalues and modes
employ finite differences or finite elements (FEM).
They handle geometric complexity well
but have two major flaws:
i) it is very cumbersome to achieve high convergence rates
and high accuracy, and ii) since
several nodes are needed per wavelength
they scale poorly as the eigenvalue $E$
grows, requiring of order $E$ degrees of freedom
(\eg for the mushroom deMenezes \etal \cite{demen07} appear limited to
$j<400$).
The numerical difficulty
is highlighted by the fact that {\em analog} computation
using microwave cavities is still popular in
awkward geometries~\cite{sridhar,dietz}.
%
%

In contrast 
we use boundary-based methods, as explained in
Sec.~\ref{sec:num}.
These i) achieve spectral accuracy, allowing eigenvalue computations
approaching machine precision
as exhibited for low-lying modes in Sec.~\ref{sec:low},
and ii) require only
of order $E^{1/2}$ degrees of freedom
(with prefactor smaller than boundary integral methods~\cite{backerbim}).
Furthermore at high $E$ we use an accelerated variant,
the scaling method~\cite{v+s,mythesis,que},
which results in another factor of order $E^{1/2}$ in efficiency.
These improvements allow us to find large numbers of
modes up to $j\sim10^5$;  
in Sec.~\ref{sec:bdry} we show such modes
along with their Husimi (microlocal) representations
on the boundary.
Visualization of modes can be an important tool,
\eg in the discovery of scars~\cite{hel84}.

We are motivated by
a growing interest in quantum ergodicity
\cite{lindenstrauss,faure}.
For purely ergodic billiards,
the Quantum Ergodicity Theorem~\cite{schnir,cdv,zel,zzw}
(QET) states that in the $E\to\infty$ limit
almost all modes become equidistributed
(in coordinate space,
and on the boundary phase space~\cite{hassell,backerhus}).
However no such theorem exists for mixed billiards,
thus numerical studies are vital.
It is a long-standing conjecture of Percival~\cite{percival}
that for mixed systems, modes
tend to localize to one or another invariant region of phase space,
with occurence in proportion to the phase space volumes,
and that those in ergodic regions are equidistributed.
(This has been tested in a smooth billiard \cite{carlo},
and recently proved for certain piecewise
linear quantum maps~\cite{marklof}).
We test the conjecture via a matrix element
\eqref{eq:foot} sensitive to the boundary
(for numerical efficiency); we then can categorize
(almost all) modes as regular or ergodic.
We address two issues which have also been raised by
recent microwave experiments in the mushroom \cite{dietz}.
i) The mechanism for dynamical tunneling~\cite{davis81}
is unknown (although it has been studied in KAM mixed billiards
\cite{fris98}).
In Sec.~\ref{sec:perc}
we propose and test a simple model for coupling strength
(related to \cite{loeck07})
which predicts observed features of matrix element distributions.
ii) The level-spacing distribution,
conjectured to be a universal feature \cite{gutz,RMT},
is studied in Sec.~\ref{sec:level}, where we also 
examine spacing distributions for regular and ergodic subsets of modes.
Note that we use an order of magnitude more modes than any existing
experiment or study.
Finally we draw conclusions in Sec.~\ref{sec:conc}.

\begin{figure}
\bc
\includegraphics[width=\wid]{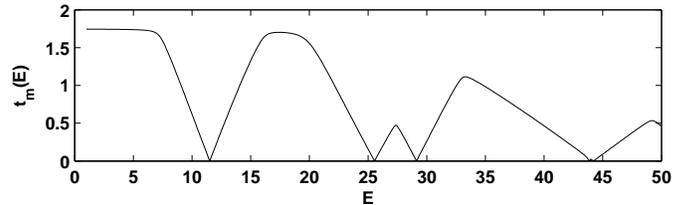}
\ec
\caption{The tension $t_m(E)$
plotted as a function of trial eigenvalue parameter $E$,
for the half-mushroom with Dirichlet boundary conditions.
The minima indicate the
eigenvalues of this domain. Close to $E=44$ there is a cluster of
two eigenvalues.
\label{fig:tmcurve}
}
\end{figure}

\section{Numerical methods}
\label{sec:num}

In this section we outline the numerical methods that make our
investigation possible; the reader purely interested in results may skip
to Sec.~\ref{sec:low}.

\subsection{The Method of Particular Solutions}

Our set of basis functions, or
{\em particular solutions}, $\{\xi_n\ofr\}_{n=1\cdots N}$
satisfy $-\Delta \xi_n = E \xi_n$
at some trial eigenvalue
parameter $E$, but do not individually satisfy (\ref{eq:bcs}). The
goal is now to find values of $E$ such that there exists 
nontrivial
linear combinations $x_1\xi_1+x_2\xi_2+\dots+x_N\xi_N$, which are small on
the boundary. These are then hopefully good approximations for an
eigenfunction.

Let us make this precise. We define the space $\mathcal{H}(E)$ of
trial functions at a given parameter $E$ as
$$
\mathcal{H}(E)=\text{Span}\{\xi_1,\dots,\xi_N\}.
$$
If we denote by $\|u\|_{\partial\Omega}$ and $\|u\|_{\Omega}$ the
standard $L^2$-norm of a trial function $u\in\mathcal{H}(E)$ on the
boundary $\pO$ and in the interior $\Omega$ we can
define the normalized boundary error (also called the tension)
as
\be
t[u]:=\frac{\|u\|_{\partial\Omega}}{\|u\|_{\Omega}}
\label{eq:tu}
\ee
It is immediately clear that $t[u]=0$ for $u\in\mathcal{H}(E)$ if
and only if $u$ is an eigenfunction and $E$ the corresponding
eigenvalue on the domain $\Omega$. However, in practice we will
rarely achieve exactly $t[u]=0$. We therefore define the smallest
achievable error as $t_m(E):=\min_{u\in\mathcal{H}(E)}t[u]$. This
value gives us directly a measure for the error of an eigenvalue
approximation $E$, namely there exists an eigenvalue $E_j$ such that
\begin{equation}
\label{eq:molerpayne}
\frac{|E-E_j|}{E_j}\leq Ct_m(E),
\end{equation}
where $C$ is an $O(1)$ constant that only depends on the domain $\Omega$. This
result is a consequence of error bounds of Moler and Payne
\cite{molerpayne,KSbounds}.
Hence, by searching in $E$ for minima of $t_m(E)$
we find approximate
eigenvalues with relative error given by a constant times $t_m(E)$.
Fig.~\ref{fig:tmcurve} shows such a plot of $t_m(E)$ for our mushroom
domain.

The implementation of this Method of Particular Solutions (MPS)
depends on i) basis set choice, and ii) how to evaluate $t_m(E)$.
The former we address in the next section.
The latter requires a set of quadrature points $\{\mbf{y}_i\}_{i=1\cdots M}$ 
on which to approximate the boundary integral $||u||_{\pO}$.
One must take into account that Helmholtz basis sets tend to be
ill-conditioned, that is, the $M\times N$ matrix $A$ with entries
$A_{in} := \xi_n(\mbf{y}_i)$ becomes numerically rank-deficient
for desirable choices of $N$.
The tension $t_m(E)$ can then be given by the square-root of the
lowest generalized eigenvalue of the matrix pair $(A^TA, B^TB)$,
or by the lowest generalized singular value of the pair $(A,B)$,
where $B$ is identical to $A$ except with the replacement of $\{\mbf{y}_i\}$
by interior points \cite{KS,mythesis,incl}.
These different approaches are discussed in \cite{gsvd}.
Here, we use the generalized singular value
implementation from \cite{gsvd}, which is highly accurate and numerically
stable.
We note that these methods are related to, but
improve upon, the plane wave method of Heller~\cite{hellerhouches}.


\subsection{Choice of basis functions}
\label{sec:bas}

In order to obtain accurate eigenvalue and eigenfunction
approximations from the MPS it is necessary to choose the right set of
basis functions. In this section we propose a basis set that leads to
exponential convergence, i.e. errors which scale as $e^{-cN}$ for some
$c>0$, as $N$ the number of basis functions grows.

To achieve this rate we first desymmetrize the problem. The
mushroom shape $\Omega$ is symmetric about a straight line going
vertically through the center of the domain (see
Fig.~\ref{fig:geom}). All eigenmodes are either odd or even
symmetric with respect to this axis. Hence, it is
sufficient to consider only the right half, $\rO$.
The odd modes are obtained by imposing zero Dirichlet boundary
conditions everywhere on the boundary $\prO$ of the half mushroom.
The even modes are obtained by imposing zero Neumann conditions on the
symmetry axis $\Gamma_s$ and zero Dirichlet conditions on the rest of
$\prO$.

Eigenfunctions of the Laplacian are analytic everywhere inside a
domain except possibly at the boundary~\cite{garab}.
Eigenfunctions can be analytically extended by
reflection at corners whose interior angle is an integer fraction
of $\pi$ \cite{KS}.
The only
singularity appears at the reentrant corner with angle $3\pi/2$
(where dashed lines meet in Fig.~\ref{fig:geom}b).
Close to this corner any eigenfunction $\phi_j$ can be
expanded into a convergent series of Fourier-Bessel functions of the
form
\begin{equation}
\label{eq:fbessel} \phi_j(r,\theta)=\sum_{n=1}^\infty
a_kJ_{\frac{2n}{3}}(k_jr)\sin\frac{2n}{3}\theta,
\end{equation}
where the polar coordinates $(r,\theta)$ are chosen as in
Fig.~\ref{fig:geom}b. The function $J_{\alpha}$ is the Bessel function
of the first kind of order $\alpha$.

The expansion (\ref{eq:fbessel}) suggests that the basis set
$\xi_n:=J_{\frac{2n}{3}}(kr)\sin\frac{2n}{3}\theta$, where $k^2=E$,
might be a good
choice since these functions capture the singularity at the
reentrant corner and automatically satisfy the zero boundary
conditions on the segments adjacent to this corner
(dashed lines in Fig.~\ref{fig:geom}b).
Hence, we only need
to minimize the error on the remaining boundary $\Gamma$ which
excludes these segments.
The boundary coordinate $q \in [0,L]$ parametrizes
$\Gamma$; its arc length is $L= 3(1 + \pi/4)$.
This Fourier-Bessel basis originates
with Fox, Henrici and Moler~\cite{fhm} for the L-shaped domain;
we believe it is new in quantum physics. In
\cite{mps} the convergence properties of this basis set are investigated
and it is shown that for modes with at most one corner
singularity the rate of convergence is exponential.
%
Indeed, in practice we find $t_m(E_1)=O(e^{-cN})$ for some $c>0$ as the number
$N$ of basis functions grows. Hence, for the minimum $\hat{E}$ of
$t_m(E)$ in an interval containing $E_1$ it follows from
(\ref{eq:molerpayne}) that
$$
\frac{|\hat{E}-E_1|}{E_1}\leq Ct_m(\hat{E})\leq
Ct_m(E_1)=O(e^{-cN}),
$$
which shows the exponential convergence of the eigenvalue
approximations $\hat{E}$ to $E_1$ for growing $N$.

\begin{figure}
a)\;\raisebox{-1.6in}{\includegraphics[width=0.95\wid]{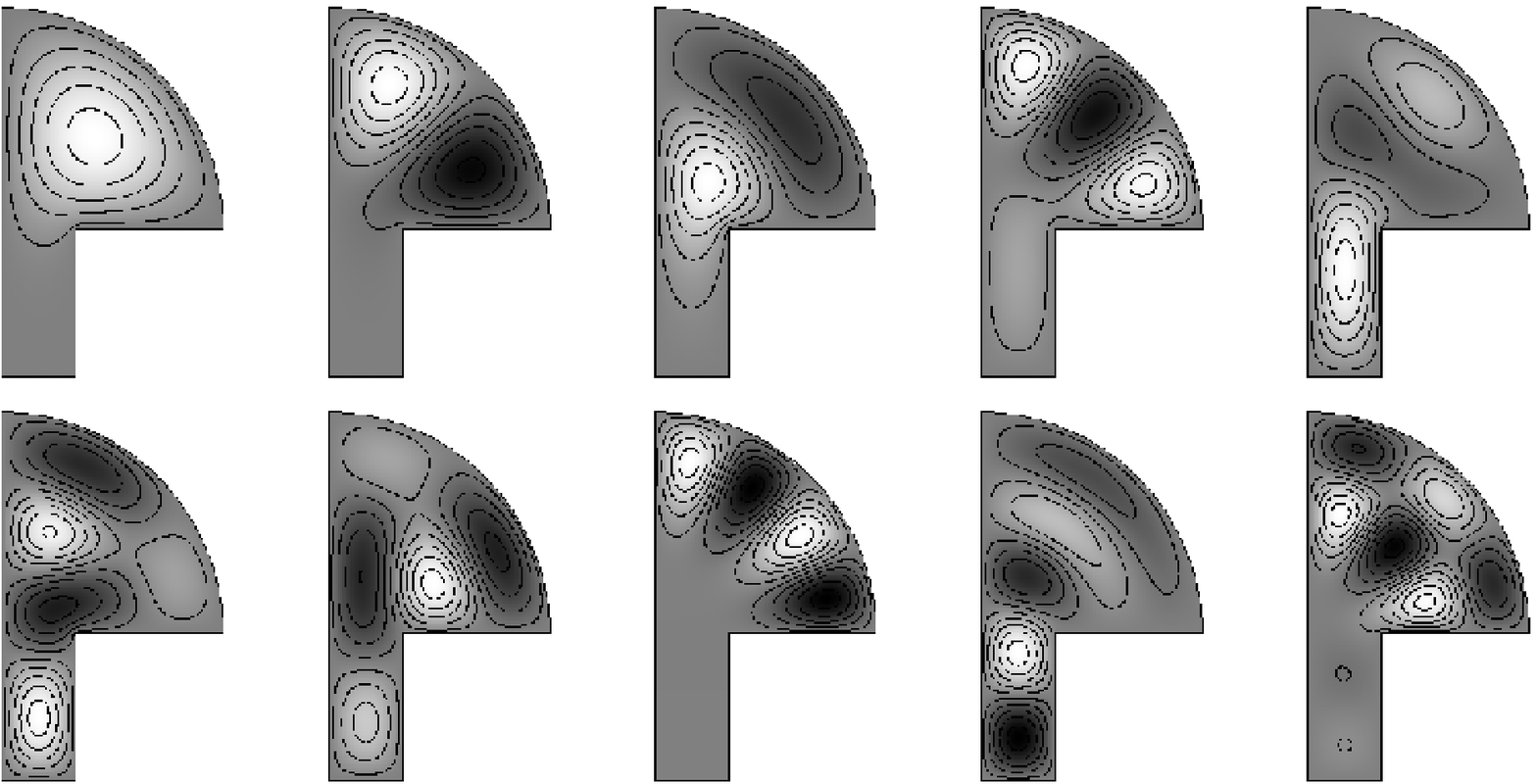}}\\
\vspace{3ex}
b)\;\raisebox{-1.6in}{\includegraphics[width=0.95\wid]{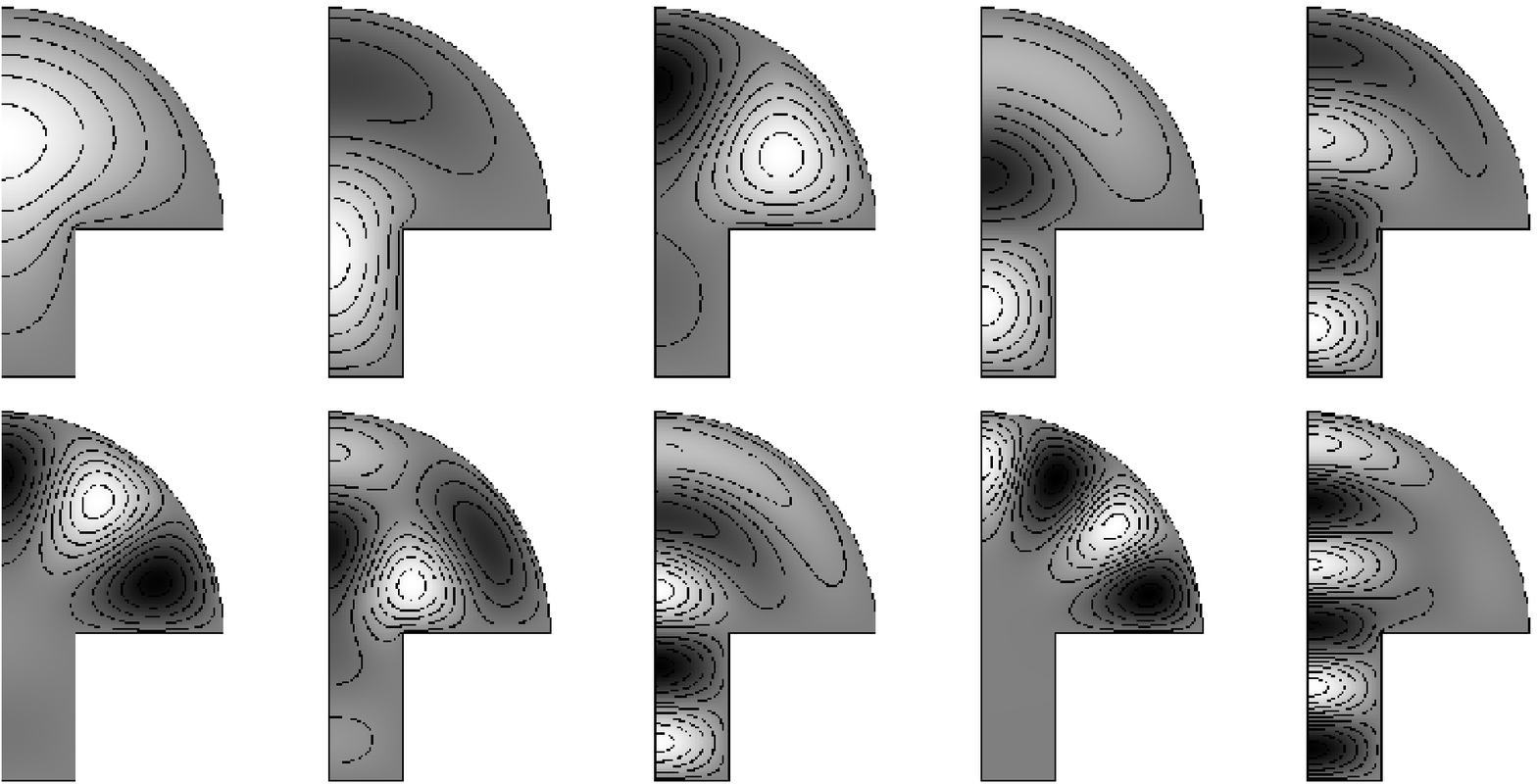}}
\caption{The first $10$ odd (a) and even (b) modes of the mushroom shape,
shown as density plots. Eigenvalue increases rightwards from
the top left.
White corresponds to positive and black to negative
values.
\label{fig:lowodd}
}
\end{figure}

\subsection{Scaling method at high eigenvalue}

For all odd modes apart from the lowest few we used an accelerated MPS variant,
the {\em scaling method} \cite{v+s,mythesis,que}, using the same basis
as above (to our knowledge the scaling method has not been combined with a
re-entrant corner-adapted basis before now).
Given a center wavenumber $k_0$ and interval half-width
$\Delta k$,
the scaling method
finds all modes $\phi_j$ with $k_j\in[k_0-\Delta k,
k_0+\Delta k]$.
This is carried out by solving
a single indefinite generalized eigenvalue problem involving
a pair of matrices of the type $A^TA$ discussed above.
The `scaling' requires a choice of origin; for technical reasons
we are forced to choose the singular corner.
Approximations to eigenvalues lying in the interval are related to
the matrix generalized eigenvalues, and the modes to the eigenvectors.
The errors grow~\cite{mythesis} as $|k_j-k_0|^3$, thus the interval width is
determined by the accuracy desired; we used $\Delta k = 0.1$
which ensured that $t_m(E)$ errors associated with the modes
rarely exceeded $3\times 10^{-4}$.
Since the search for minima required by the MPS has been avoided,
and on average $O(k)$ modes live in each interval,
efficiency per mode is thus $O(k)=O(E^{1/2})$ greater than the MPS.
By choosing a sequence of center wavenumbers $k_0$
separated by $2\Delta k$,
all modes in a large interval may be computed.
Rather than determine the basis size $N$ by a convergence criterion
as in Sec.~\ref{sec:bas}, for $E>10^3$ we use the Bessel function 
asymptotics:
for large order $J_\alpha(x)$ becomes exponentially
small for $x/\alpha < 1$ (the turning point is $x=\alpha$).
Equating the largest argument $kR$ (with $R=3/2$) with the 
largest order $2N/3$ gives our semiclassical basis size
$N \approx 9k/4 = O(E^{1/2})$.

\begin{figure}[h]
\bc
\includegraphics[width=\wid]{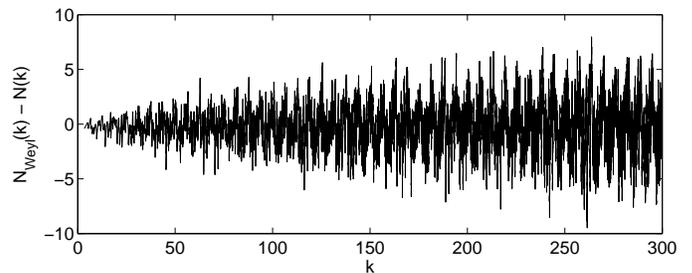}
\ec
\caption{Difference between the mode counting function $N(k)$
and the two-term Weyl's prediction $N_\tbox{Weyl}(k)$ defined
by (\ref{eq:weyl}), for the 16061 odd-symmetric eigenvalues with $k_j<300$.
\label{fig:weyl}
}
\end{figure}

We are confident that the
scaling method finds all odd modes in a desired eigenvalue window.
For instance we compute all 16061 odd symmetry modes with $k_j<300$,
using 1500 applications of the scaling method (at
$k_0 = 0.1, 0.3, \ldots ,299.9$).
This computation takes roughly 2 days of CPU time
\footnote{All calculation times are reported for one core of a 
2.4 GHz Opteron running C++ or MATLAB under linux/GNU}.
We verify in Fig.~\ref{fig:weyl}
that there is zero mean fluctuation in the
difference between the 
(odd) level-counting function $N(k) :=\#\{j:k_j\le k\}$
and the first two terms of Weyl's law~\cite{gutz},
\be
N_\tbox{Weyl}(k) = \frac{\vol(\rO)}{4\pi} k^2 - \frac{|\prO|}{4\pi} k,
\label{eq:weyl}
\ee
where $|\prO|$ is the full perimeter of the half mushroom domain.
Note that there is no known variant of the
scaling method that can handle Neumann or mixed BCs,
hence we are restricted to odd modes.
It is interesting that the
method is still not completely understood from the numerical
analysis standpoint~\cite{v+s,mythesis,que}.

In applying the scaling method to the mushroom, the vast majority of
computation time involves evaluating Bessel functions $J_\alpha(x)$
for large non-integral $\alpha$ and large $x$.
This is especially true for producing 2D
spatial plots of modes as in Fig.~\ref{fig:high}, for which of order
$10^9$ evaluations are needed (1 hr CPU time).
We currently
use independent calls to the GSL library~\cite{GSL} for each $J_\alpha(x)$
evaluation.
This is quite slow, taking between 0.2 and 50 $\mu$s per call,
with the slowest being in the region $\alpha<50$, $10^2<x<10^3$.
However, we note that Steed's method~\cite{ross,numrec}, which is what
GSL uses in this slow region, is especially fast at evaluating
sequences $J_\alpha(x), J_{\alpha-1}(x), J_{\alpha-2}(x),\ldots$,
and that since $\alpha$ is a multiple of a rational with denominator 3,
only 3 such sequences
would be needed to evaluate all basis functions
$\{\xi_m\ofr\}_{m=1\cdots M}$ at a given location $\mbf{r}$.
We anticipate at least an
order of magnitude speed gain could be achieved this way.

\begin{figure}[ht]
\bc
\includegraphics[width=\wid]{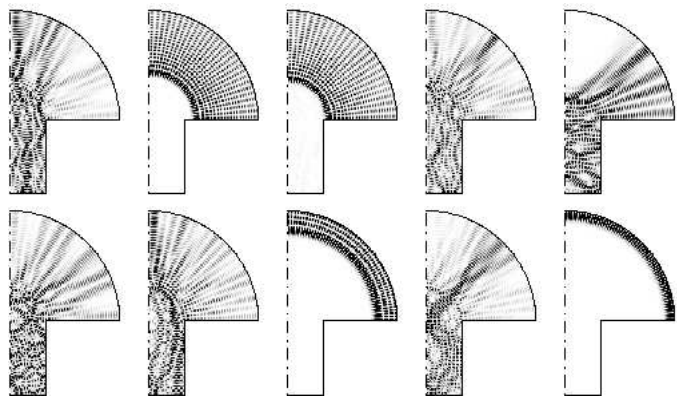}
\ec
\caption{The 10 odd modes of the mushroom whose eigenwavenumbers
lie in the range $90<k_j<90.35$, at mode number about $j\approx 1430$.
Intensity $|\phi_j|^2$ is shown with zero white and larger values darker.
\label{fig:medi}
}
\end{figure}

\begin{table} 
\begin{center}
a)
\raisebox{-.9in}{
\begin{tabular}{c|l}
$j$ & $E_j$\\
\hline
1 & 11.50790898\\ 2 & 25.55015254\\
3&29.12467610 \\4 &43.85698300\\
5&44.20899253  \\ 6&53.05259777\\
7&55.20011630 \\8&66.42332921\\
9&69.22576822 \\10&82.01093712
\end{tabular}
}
\qquad\qquad
b)
\raisebox{-.9in}{
\begin{tabular}{c|l}
$j$ & $E_j$\\
\hline
1&5.497868889 \\2&13.36396253\\
3&18.06778679  \\4&20.80579368\\
5&32.58992604  \\6& 34.19488964\\
7&41.91198264 \\8&47.37567140\\
9&54.62497098  \\10&65.18713235
\end{tabular}
}
\end{center}
\caption{Tables of a) lowest 10 odd and b) lowest 10 even
eigenvalues of the mushroom.
All digits shown are believed to be correct.
\label{tbl}
}
\end{table} 

\section{Low eigenvalue modes}
\label{sec:low}

In this section we present highly-accurate results for the first few
even and odd modes.
Odd modes are
obtained by solving the eigenvalue problem with zero Dirichlet
boundary conditions on the half mushroom from
Fig.~\ref{fig:geom}b, using the MPS, by locating
minima in the tension function of Fig.~\ref{fig:tmcurve}.
In Table~\ref{tbl}a the eigenvalues are listed to at least 10 significant
digits, and in Fig.~\ref{fig:lowodd}a the corresponding modes are plotted.
We emphasize that it is the exponential convergence of our basis
that makes such high accuracies a simple task.

For even modes we impose
Neumann BCs on $\Gamma_s$ and Dirichlet BCs on the remaining
part of $\Gamma$. This was achieved in the MPS
by modifying the tension function (\ref{eq:tu}) to
read
\be
t[u]:=\frac{\left(\|\partial_n
u\|_{\Gamma_s}^2+\|u\|_{\Gamma\backslash\Gamma_s}^2\right)^{1/2}}
{\|u\|_{\Omega'}}
\label{eq:tueven}
\ee
where the normal derivative operator on the boundary is
$\partial_n:=\mbf{n}\cdot\nabla$, the unit normal vector being $\mbf{n}$.
Table~\ref{tbl}b, gives the smallest $10$
even modes on the mushroom billiard,
and the corresponding modes are plotted in Fig.~\ref{fig:lowodd}b.

Although we are far below the semiclassical regime we already see
properties of the underlying
classical dynamical system. For example, the $8^{th}$ odd and the $6^{th}$
even mode live along a caustic and therefore show features of the
classically integrable phase space while the $7^{th}$ odd
and $10^{th}$ even mode already
shows features of the classically ergodic phase space.
For comparison, in Fig.~\ref{fig:medi} we show some odd
modes with intermediate 
eigenvalues of order $10^4$ (odd mode number of order $10^3$),
a similar quantum number to that measured in a microwave cavity
by Dietz \etal \cite{dietz}.
As these authors noted, modes
at this energy usually live in either the integrable or
to the ergodic regions of phase space;
we pursue this in detail in Sec.~\ref{sec:perc}.

\begin{figure}[h]
\bc
\includegraphics[width=\wid]{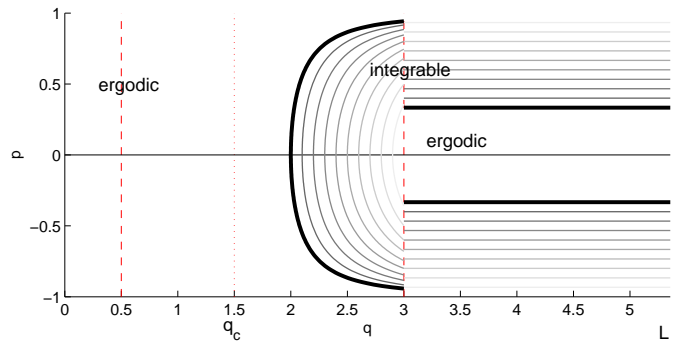}
\ec
\caption{\poin surface of section (PSOS), that is, the classical phase space
in boundary coordinates $q$ (as shown in Fig.~\ref{fig:geom}b) and
$p$ (sin of incidence angle).
Vertical dashed lines shows location of corners.
The vertical dotted line shows location of $q_c$ the focal point of the hat.
The dark line shows the border of integrable phase space; note that
$q=2$ corresponds to the smallest possible caustic for integrable phase space.
Families of orbits defined by constant angular momentum are shown by
lines in the integrable region.
Note that they exchange vertical ordering at the corner, as indicated by
their grayscale color labeling.
\label{fig:psos}
}
\end{figure}

\begin{figure}[h]
\bc
\includegraphics[width=\wid]{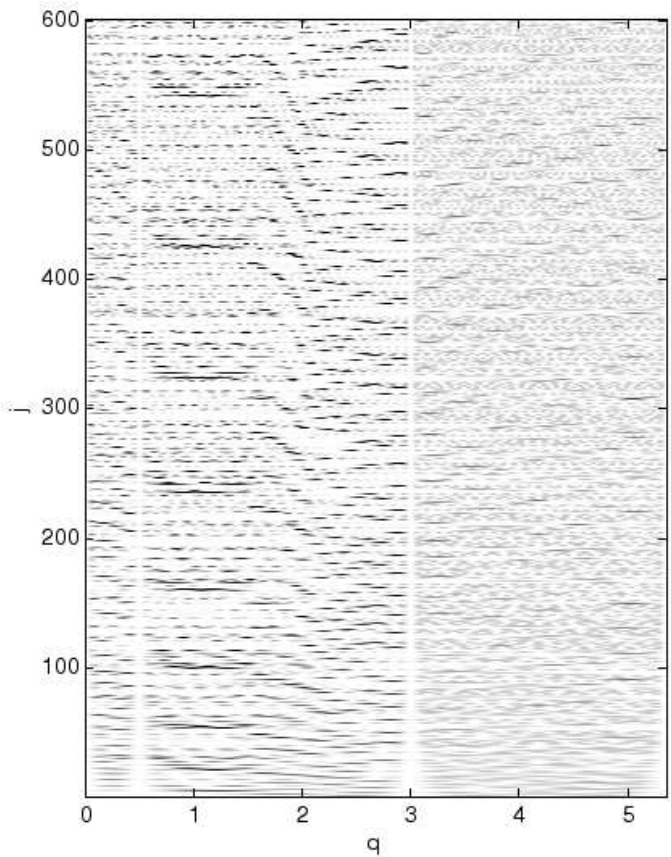}
\ec
\caption{
Intensity of boundary normal-derivative functions
$|\nd{j}(q)/k_j|^2$,
plotted vs boundary coordinate
$q$ on the horizontal axis and odd mode number $j\in[1,600]$ on the vertical.
The density plot shows white as zero, and larger values darker.
\label{fig:ngr}
}
\end{figure}

\section{Boundary and Husimi functions}
\label{sec:bdry}

We choose a \poin
surface of section (PSOS)~\cite{gutz}
defined by Birkhoff coordinates $(q,p) \in \Gamma \times [-1,1] =: Z$,
where $q$ is the boundary location as before (see Fig.~\ref{fig:geom}b)
and $p$ the tangential velocity component, in the clockwise sense,
for a unit speed particle. (If the incident angle from the normal
is $\theta$ then $p=\sin \theta$).
The structure of this PSOS phase space is shown in Fig.~\ref{fig:psos}.
Our choice (which differs from that of Porter \etal \cite{porter})
is numerically convenient since it involves only
the part of the boundary on which matching is done (Sec.~\ref{sec:num}).
Despite the fact that it does not cover the whole
boundary $\prO$, it is a valid PSOS since all trajectories must hit
$\Gamma$ within bounded time.

Integrable phase space consists of precisely the orbits which, for all time,
remain in
the hat~\cite{mushroom}
but which never come within a distance $b/2$ from the center point $q_c$
\footnote{%
This requirement is needed to exclude the zero-measure set of
marginally-unstable periodic orbits
(MUPOs) in the ergodic region which nevertheless remain in the the hat
for all time~\cite{sticky1,sticky2}}.
Simple geometry shows that
the curved boundary between ergodic and integrable regions
consists of points $(q,p)$ satisfying
\be
q - q_c \; = \; \frac{b/2}{\sqrt{1 - p^2}},\quad \mbox{for}\;\;
 p^2\le p_0^2 := 1 - \frac{b^2}{4R^2}.
\ee
For our shape, $q_c = a + b/2 = 3/2$, $p_0^2 = 8/9$.
In the domain $q\in[q_c+R,L]$ the boundary occurs at the lines $p=\pm b/2R
= \pm 1/3$. 
Successive bounces that occur on $\Gamma$
are described by the PSOS billiard map $f:Z\to Z$.
Any such \poin map is symplectic and therefore area-preserving~\cite{gutz}.

The quantum boundary functions $\nd{j}(q)$
for $q\in[0,L]$
are convenient and natural representations of the modes.
Note that they are not $L^2(\pO)$ normalized;
rather they are normalized according to a geometrically-weighted
$L^2$ boundary norm
via the Rellich formula (see \cite{rellich,que})
\be
\int_\pO (\mbf{r}\cdot\mbf{n})\, |\nd{j}|^2\, dq = 2E_j ,
\label{eq:rellich}
\ee
where $\mbf{r}(q)$ is the location of boundary point $q$ relative to
an arbitrary fixed origin.
Fig.~\ref{fig:ngr} shows the
intensities of the first 600 odd boundary functions.
Features include
an absence of intensity near the corners (over a region whose
size scales as the wavelength).
The region $3<q<L$, in which phase space is predominantly integrable,
has a more uniform intensity than $0<q<2$, which is exclusively ergodic.
The region $2<q<3$ is almost exclusively integrable, but is
dominated by classical turning-points corresponding to caustics;
these appear as dark Airy-like spots.
In $1/2 < q< 3/2$ there are horizontal dark streaks corresponding
to horizontal `bouncing-ball' (BB)
modes
in the foot.
Finally, a series of slanted dark streaks is visible for $3/2<q<2$:
these interesting fringes move as a function of wavenumber and
we postpone analysis to a future publication.

\begin{figure}[ht]
\bc
\includegraphics[width=\wid]{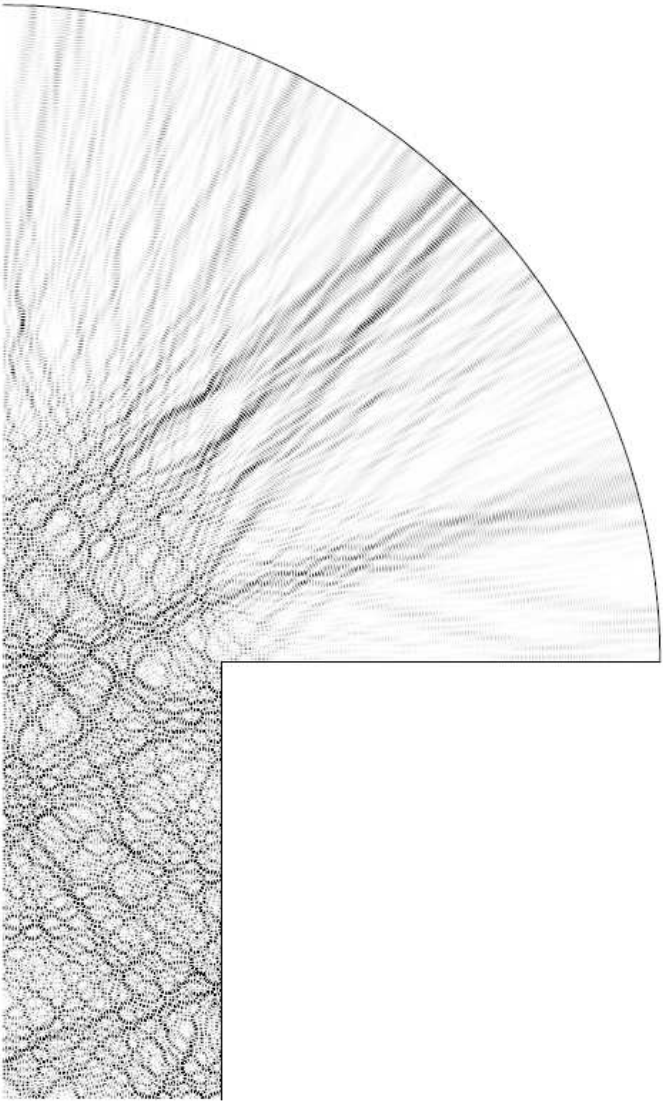}
\ec
\caption{
High-energy eigenmode with $k_j=499.856\cdots$, at around odd mode number
$j\approx 45000$. This mode appears to live in the ergodic region.
\label{fig:high}
}
\end{figure}

\begin{figure}[t]
\bc
\includegraphics[width=0.8\wid]{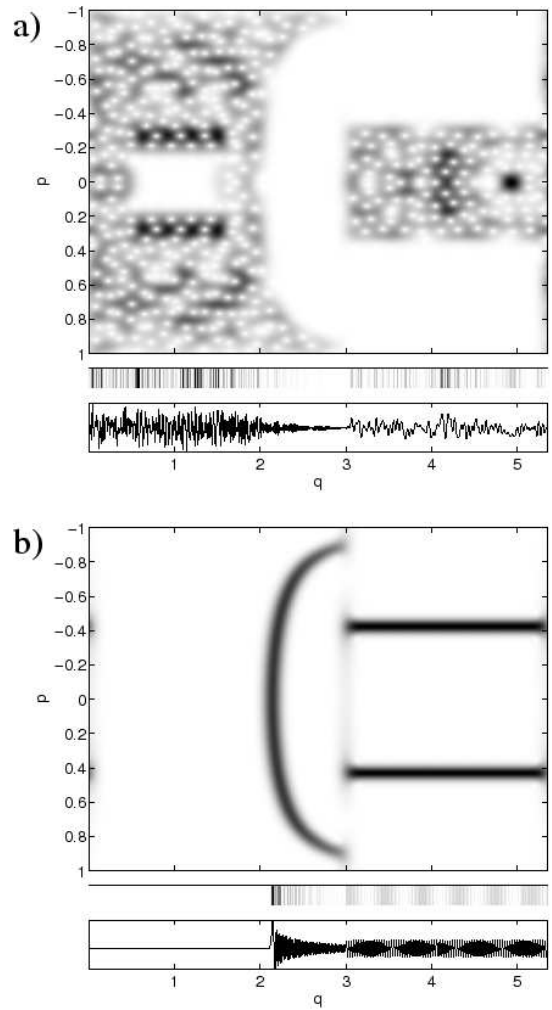}
\ec
\caption{
a) the mode of Fig.~\ref{fig:high}
Husimi distribution $H_{\nd{j},\sigma}(q,p)$ defined by (\ref{eq:hus}) (top),
density plot of $|\nd{j}|^2$ (middle),
and graph of $\nd{j}$ (bottom). Note the $q$ coordinate is common to the
three plots.
b)  Similar representation of the next highest mode at $k_j=499.858$,
the 15$^{th}$ in the sequence of Fig.~\ref{fig:many},
which lives in the regular region.
\label{fig:hus}
}
\end{figure}


In Fig.~\ref{fig:many} we show a sequence of 20 much higher
modes with consecutive
eigenvalues near wavenumber $k=500$ (eigenvalue $E=2.5\times 10^5$).
These modes are a subset of the modes
produced via a single generalized
matrix eigenvalue problem (of size $N \approx1200$)
using the scaling method at $k_0=500$.
The full set of 77 modes (evaluating boundary functions)
took only 20 mins CPU time.
Typical tension $t_m(E)$ values were below
$10^{-3}$. Naively applying (\ref{eq:molerpayne}) we would conclude
only about 3 relative digits of accuracy on eigenvalues.
However, it is possible to rigorously improve this bound by
factor $O(E_j^{1/2})$ \cite{incl}, giving about 6 digits.

Fig.~\ref{fig:high} shows the 14th in the sequence in more detail.
The corresponding boundary function
is shown in Fig.~\ref{fig:hus}a,
along with the intensity, and its Husimi distribution.
The Husimi distribution is a coherent-state
projection of the mode onto the
PSOS phase space (see App.~\ref{app:hus}).
The choice of the aspect ratio $\sigma$
is somewhat arbitrary but it is expected~\cite{hel84}
that phase space structures have spatial scale $O(k^{-1/2})$,
so we chose a scaling similar to this: with $k=500$ we used $\sigma=0.076$.
By comparing to the phase space (Fig.~\ref{fig:psos})
we see localization to the ergodic region.
The only part of ergodic phase space not well covered
contains BB modes in the foot (the white `box').
A scar is also visible as the 9 darkest spots: 4 pairs of spots
surrounding the white box correspond to 4
bounces in the foot, and a single spot at $q\approx 5$ corresponds to
a normal-incidence bounce off the circular arc.
By contrast, Fig.~\ref{fig:hus}b
shows the boundary function of a mode living in the regular region
(the 15th in Fig.~\ref{fig:high}); the energy-shell localization is clear.
The full set of 20 Husimi functions is shown
in Fig.~\ref{fig:manyhus}.
We remind the reader that in purely ergodic systems boundary functions obey the
QET~\cite{hassell,backerhus} with almost every $\nd{j}/k_j$ tending to
an invariant Husimi density of the form $C\sqrt{1-p^2}$.
We might expect a similar result for the ergodic subset of modes in the
ergodic phase space of the mushroom.
However, Fig.~\ref{fig:manyhus} highlights that, despite
being at a high mode number of roughly $4\times 10^4$,
we are still a long way from reaching
any invariant density: the 7 ergodic modes have highly non-uniform
distributions.

\begin{figure}[ht]
\bc
\includegraphics[width=\wid]{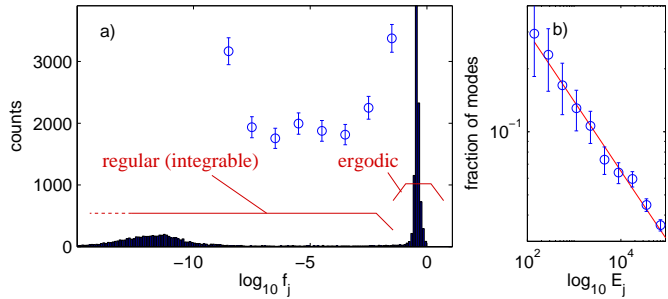}
\ec
\caption{
a) Histogram of the logarithm of
$f_j$, the `foot-sensing' matrix element \eqref{eq:foot},
for the first 16061 odd modes. 
Errorbars show counts of $f_j$ lying in each decade
(errors assuming independent counts), on a vertical
scale magnified by a factor 15.
b) Fraction of modes with $10^{-8}<f_j<10^{-2}$
lying in logarithmically-spaced $E_j$ intervals
(errorbars), compared to power law $E^{-1/3}$ (solid line).
\label{fig:fjhist}
}
\end{figure}

\section{Percival's conjecture and dynamical tunneling}
\label{sec:perc}

In the small set of 20 high-lying modes discussed above,
Percival's conjecture holds:
modes are either regular or chaotic but not a mixture of both.
We will now study this statistically with a much larger set,
the first $n=16061$ odd modes corresponding to $0<k_j<300$.
Since the PSOS phase space in $0<q<3/2$ is ergodic for all $p$,
the following 
`foot-sensing' quadratic form, or diagonal matrix element,
is a good indicator of an ergodic component:
\be
f_j := \frac{1}{2E_j}\int_0^{3/2} (\mbf{r}\cdot\mbf{n})
\left|\nd{j}\right|^2 dq,
\label{eq:foot}
\ee
where, as Fig.~\ref{fig:geom}b shows,
$\mbf{r}\cdot\mbf{n}$ takes the value 1 for $0<q<1/2$, and
$1/2$ for $1/2<q<3/2$.
(The weighting by $\mbf{r}\cdot\mbf{n}$ is chosen to mirror
\eqref{eq:rellich};
scaling by $E_j$ is necessary for a well-defined semiclassical limit
\cite{hassell,backerbdry}).

The observed distribution of $f_j$ is shown in Fig.~\ref{fig:fjhist}a.
The main
feature is a cluster around $O(1)$ (we associate with ergodic modes)
and a wider distribution of smaller values (predominantly regular modes).
We have tested that the
apparent cluster lying roughly from $10^{-14}$ to $10^{-9}$
is merely an artifact reflecting the size of
numerical errors in $\nd{j}$:
the key point is that there is a {\em continuum} of values
(see errorbars in Fig.~\ref{fig:fjhist}a) which extends from $O(1)$
down to exponentially small values.
Roughly 0.75\% of the total number of modes fall within each decade
from $10^{-2}$ to $10^{-8}$.
We believe that in the absence of numerical errors a similar distribution would
extend down many tens of orders of magnitude.

Percival's conjecture would imply that
the sequence $\{f_j\}_{j=1\cdots\infty}$
has (for all but a set of vanishing measure) two limit points:
zero (for regular modes), and some positive constant (for ergodic modes).
Even though most mode numbers are large ($\sim10^4$) the upper cluster
still has a wide standard deviation of 0.1 (its mean is 0.39); this
is in line with our recent work
confirming the slow algebraic semiclassical convergence of matrix elements
\cite{que}.

We would like to test whether the relative mode frequencies
of regular vs ergodic modes are in
proportion to the corresponding classical phase-space volumes.
We categorize modes by defining them as `regular' if $f_j<0.1$.
This choice of cutoff value is necessarily a compromise between lying
below the whole ergodic peak yet capturing the full dynamic range of
regular modes.
This gives a fraction
$\alpha_\tbox{reg} := n_\tbox{reg}/n = 7178/16061 = 0.4469\cdots$
of regular modes,
which is only 1.7\% less than the integrable phase space fraction
$\mu_\tbox{reg} =  0.4549\cdots$ (computed in App.~\ref{app:uint}).
Assuming that each regular mode counted arose randomly and
independently due to some underlying rate (fraction of level density),
we may associate a standard error of
$\sqrt{n_\tbox{reg}(n-n_\tbox{reg})/n^3} = 0.004$
with the measured fraction.
Thus the discrepancy is only 2 sigma, not inconsistent with the (null)
hypothesis that $\alpha_\tbox{reg} = \mu_\tbox{reg}$.
To check whether this result persists semiclassically
we computed a smaller set of $n=615$ high-lying modes
sampled from the range $500<k_j<750$, up to mode number
$j\approx 10^5$, and found
$\alpha_\tbox{reg} = 0.441\pm 0.015$, again consistent with
Percival's conjecture.

\begin{figure}[ht]
\bc
\includegraphics[width=\wid]{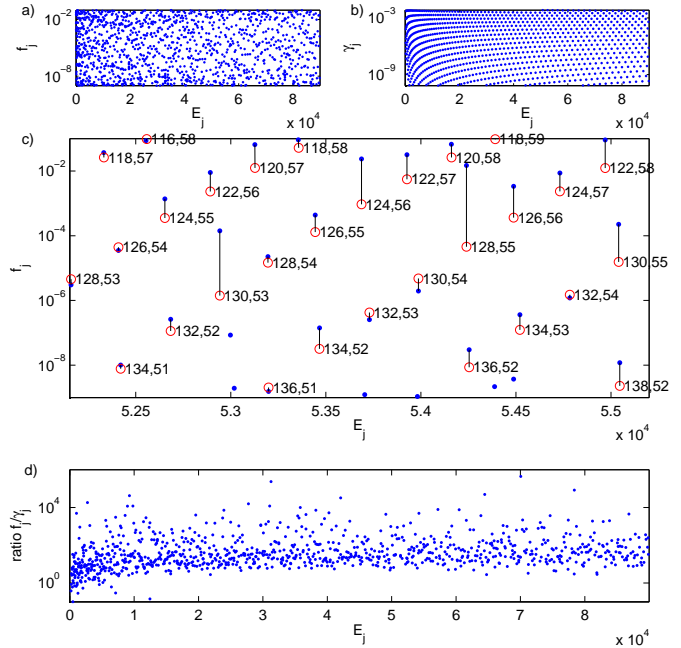}
\ec
\caption{
a) Distribution of matrix elements $f_j$
for regular modes, as function of eigenvalue $E_j$.
b) Corresponding rates $\gamma_j$ predicted by the model in the text.
c) Zoom of a), showing $f_j$ (dots) connected by vertical
lines to best-matching values of $c\gamma_j$ (circles
labeled by disc quantum numbers $m,n$), using constant $c=15$.
d) ratio $f_j/\gamma_j$ for all matched pairs.
\label{fig:fjmatch}
}
\end{figure}

\subsection{Results and model for dynamical tunneling}

The continuum of matrix element values in Fig.~\ref{fig:fjhist}a is a
manifestation of dynamical tunneling \cite{davis81}, quantum
coupling between regular and ergodic invariant phase space regions.
This has recently been seen in mushroom microwave cavity
modes~\cite{dietz}, and these authors raised the question
as to the mechanism for tunneling in this shape.
We address this by proposing and numerically testing a simple such model.
First we notice that the density of $\ln f_j$ is roughly constant
(in the range $f_h>10^{-8}$ where numerical errors are negligible).
This suggests a coupling strength which is the {\em exponential}
of some uniformly-distributed quantity.
We may ask whether this density is dependent on
eigenvalue magnitude (energy): Fig.~\ref{fig:fjhist}b shows that the
density appears to die as $E^{-1/3}$,
consistent with the expectation that all $f_j$ values for regular modes
vanish in the semiclassical limit.

Our model is to assume that $f_j$ values are controlled by a
matrix element $\gamma_j$ giving the {\em rate} of dynamical tunneling from the
regular to the ergodic region.
Each regular mode closely approximates an $(n,m)$-mode of the quarter disc,
which are the product of angular function $(2/\sqrt{\pi})\sin m\theta$
and radial function
\be
\psi_{mn}(r) = \frac{\sqrt{2}}{R J'_m(k_{mn} R)} J_m(k_{mn}r)
\ee
where $n=1,2,\ldots$ is the radial mode number and $m=2,4,\ldots$ the
angular mode number, and $k_{mn}R$ is the argument of the
$n^{th}$ zero of the $J_m$ Bessel function. Quarter-disc
eigenwavenumbers are $k_{mn}$.
The normalization is $\int_0^R |\psi_{mn}(r)|^2 \,rdr = 1$.
A wavepacket initially launched from such a disc mode will, in the mushroom,
leak into the ergodic region due to the openness of the
connection into the foot.
We take the rate proportional to the probability mass of $\psi_{mn}$
`colliding' with the foot,
\begin{gather}
\gamma_j := \int_0^{b/2} |\psi_{mn}(r)|^2 \,rdr
\label{eq:gj}
\\
= \frac{4}{[k_{mn} RJ'_m(k_{mn}R)]^2} \sum_{l=0}^\infty
(m+1+2l) \left|J_{m+1+2l} \left(\frac{b}{2}k_{mn}\right)\!\right|^2
\nonumber
\end{gather}
where we used \cite[Eq.~11.3.2]{a+s} to rewrite the integral.
This model is similar to that proposed
recently by B\"{a}cker \etal
\cite{loeck07} (in our case the `fictitious integrable system' is the
quarter-disc).
$\gamma_j$ is exponentially small only when the Bessel function turning point
lies at radius greater than $b/2$;
at eigenvalue $E$ this occurs for $b\sqrt{E}/(2m)<1$.

We compare in Fig.~\ref{fig:fjmatch}a) and b) $f_j$ values for regular
against $\gamma_j$ values computed using all relevant
$(m,n)$ quantum numbers for the quarter-disc.
It is clear that although the densities are similar, $f_j$ is irregularly
distributed whereas $\gamma_j$ values fall on a regular lattice.
However, upon closer examination there is a strong correlation.
We attempted to match each disc mode $(m,n)$ seen in panel b) with
its corresponding mushroom mode $j$ as seen in panel a);
in most of the 1051 cases there was a very clear match, with
relative eigenvalue
difference $|E_j-k_{mn}^2|/E_j < 10^{-4}$ in 90\% of the cases, and
$|E_j-k_{mn}^2|/E_j < 3\times10^{-6}$ in 74\% of cases.
(Note that, although it is not needed for our study, it
would likely be possible to improve the fraction matched using data
from $\nd{j}$.)
As shown in panel c), $f_j$ values are quite correlated with the
$\gamma_j$ values of their matched mode.
Note that an overall prefactor of $c=15$ was included to improve the fit.
The resulting ratio $f_j/\gamma_j$ is shown in panel d),
and has a spread of typically a factor $10^2$.
Since this is much less than the spread of $10^8$ in the original matrix
elements, this indicates that the above model is strongly predictive
of dynamical tunneling strength, mode for mode.
We suggest the remaining variation, and the value of $c$, might be explained by
varying eigenvalue gaps (resonant tunneling)
between quarter-disc and ergodic modes
(such variation is discussed in~\cite{fris98}), although this is an open
question.
Also in this simple model it is clear from the $E$-dependence in panel d) that
there are algebraic prefactors that should be included in a more
detailed model.

Using the model we may predict the decay $\sim E^{-1/3}$ in the density of
$f_j$ values
reported above, by returning to the sum in \eqref{eq:gj}.
For regular modes where $\gamma_j \ll 1$,
the Bessel functions in the $l\ge1$ terms
have turning points successively further away from $b/2$, thus 
the sum may be approximated by the $l=0$ term
(this has been checked numerically).
We make the approximation that the turning point is close to $b/2$,
that is $\eps\ll1$, where
\be
\eps \;:=\; 1 - \frac{bk_{mn}}{2m} \quad .
\label{eq:eps}
\ee
We focus on the exponentially small behavior of $\gamma_j$
and drop algebraic prefactors.
In \eqref{eq:gj}
using Debye's asymptotics for the Bessel function \cite[Eq. 9.3.7]{a+s}
and keeping leading terms for small $\eps>0$ gives
\be
g:=-\ln \gamma_j \;\approx\; (\mbox{const}) + \frac{1}{2}\ln \eps + \ln k_{mn}
+\frac{4\sqrt{2}}{3}m \eps^{3/2}.
\label{eq:lngj}
\ee
(This can be interpreted as the tail of the Airy approximation to the
Bessel).
For fixed $\eps\ll1$ we need keep only the last term as $m\to\infty$.
Fixing $m$ while increasing $n$ by 1
causes a small wavenumber change
$k_{m,n+1}-k_{mn} \approx \pi/(p_0 R)$, causing via \eqref{eq:eps} a change
$\Delta\eps \approx -\pi b/(2mp_0R)$, which in turn causes via \eqref{eq:lngj}
a change
\be
\Delta g \;\approx\; -\pi \sqrt{2\eps}/(p_0 R)
\;\approx\; -\frac{\pi}{p_0 R} \left(\frac{3g}{2m}\right)^{1/3}~,
\ee
where in the last step we expressed $\eps$ in terms of the asymptotic for $g$.
Realising that, for $\eps\ll 1$ we have $m\approx bk_{mn}/2 = b\sqrt{E}/2$,
and that adjacent curves of constant $m$ in the $(E,g)$-plane are separated
in $E$ by $\Delta E \approx 8\sqrt{E}/b$,
gives our result, the density of points in the $(E,g)$-plane,
\be
d(E,g) \;= \;\frac{1}{|\Delta g \,\Delta E|} \;\approx\;
\frac{p_0R}{8\pi}\left(\frac{b^4}{3 gE}\right)^{1/3} \quad .
\label{eq:dEg}
\ee
Recall that Fig.~\ref{fig:fjmatch}a,b,c illustrate the $(E,g)$-plane.
In Fig.~\ref{fig:fjhist}a
small dynamic range and counting statistics prevents this
weak dependence of density on $g$ from being detected.
However the main conclusion from \eqref{eq:dEg} is that
the density of $\gamma_j$ (and hence $f_j$) values lying in any fixed interval
scales asymptotically as $E^{-1/3}$, in agreement with Fig.~\ref{fig:fjhist}b.






\begin{figure}[ht]
\bc
\includegraphics[width=\wid]{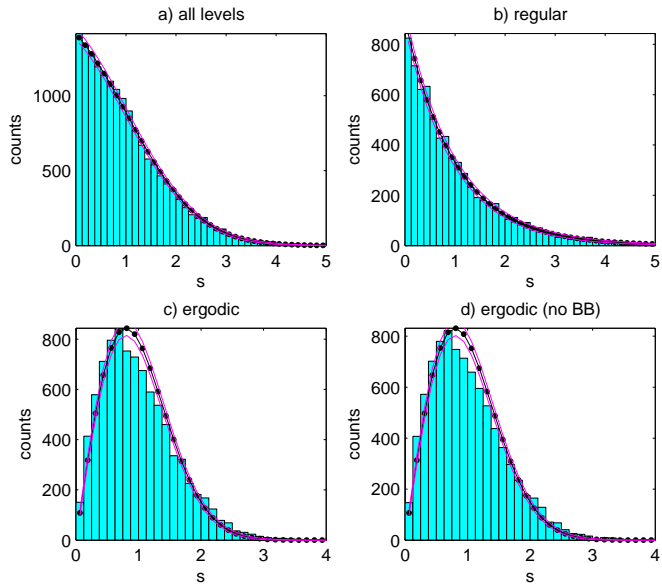}
\ec
\caption{
Nearest-neighbor spacing distributions (NNDS) $p(s)$
for the 16061 modes with $k_j<300$,
estimated via a histogram with bins of width $\Delta s = 0.125$.
Data (in counts per bin) are shown by bars.
Predictions are shown by dots, with $\pm 1$
standard error (solid lines above and below).
a) all modes, vs Berry-Robnik formula,
b) regular modes, vs Poissonian formula $e^{-s}$,
c) ergodic modes including BB modes, vs Wigner's
approximate GOE formula $\frac{\pi}{2}se^{-\pi s^2/4}$, and
d) ergodic modes with BB modes removed using the categorization in
Sec.~\ref{sec:level}, vs the same.
\label{fig:level}
}
\end{figure}

\begin{figure}[ht]
\bc
\includegraphics[width=\wid]{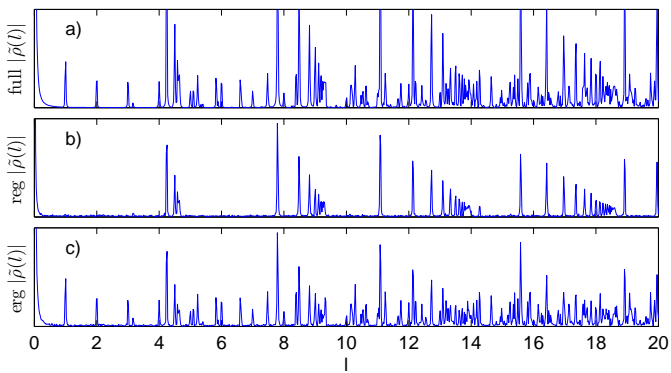}
\ec
\caption{
Absolute value of the
Fourier transform $\tilde{\rho}(l)$ of the density of states
$\rho(k):=\sum_{j=1}^\infty \delta(k-k_j)$, vs orbit length $l$,
for a) all eigenvalues lying below $k_j<300$, b) regular modes only,
c) ergodic modes only.
\label{fig:dos}
}
\end{figure}

\section{Level spacing distribution and level density fluctuation}
\label{sec:level}

We show the nearest-neighbor spacing distribution (NNDS)
of the complete set of the first $n=16061$ eigenvalues of odd-symmetric
modes $E_j$ with $k_j<300$,
in Fig.~\ref{fig:level}a.
Spacings were unfolded in the standard way \cite{RMT}, thus a histogram of
$s_j:=(E_{j+1}-E_j)/\overline{E}$, where $\overline{E}$ is the mean
level spacing, was collected.
This is compared in the figure against the Berry-Robnik
prediction~\cite{BRdist} for a mixed system with a single regular
component (of phase-space fraction $\mu_\tbox{reg} = 0.4549\cdots$)
and single ergodic component.
The agreement is excellent, with deviations consistent with
the standard error for each bin count.
In their recent work Dietz \etal \cite{dietz}
claim that there is a dip in the NNDS around $s=0.7$ associated with
supershell structure in the hat (two periodic orbits of close lengths).
Their choice of mushroom shape differs from ours only in the foot.
Our results, computed using over 16 times their number of levels,
show no such dip.
This 
suggests that their observed dip is a statistical
anomaly, or that it does not carry over to the rectangular-foot mushroom
and therefore is not associated with the hat.

In order to study this further we computed the partial NNDS
associated with regular or ergodic modes, categorized using the method
of Sec.~\ref{sec:perc}.
Regular modes (Fig.~\ref{fig:level}b) fit the Poisson level spacing
distribution well.
Ergodic modes (Fig.~\ref{fig:level}c) fit Wigner's standard approximate
form for the GOE distribution reasonably well,
however there are visible deviations:
the data systematically favors small spacings $s<0.75$ while
disfavoring intermediate spacings $0.75<s<1.6$.
This can be quantified by comparing 0.392,
the fraction of spacings with $s<0.75$, to 0.357, the corresponding
fraction predicted using the Wigner distribution.
Using the normal approximation to the 
binomial distribution, this discrepancy is nearly 7$\sigma$ and is thus
statistically very significant (similar conclusions are reached
by the standard Kolmogorov-Smirnov test for comparing distributions).
We conjecture that, as with mode intensities discussed above,
the discrepancy is another
manifestation of slow convergence to the semiclassical limit.

One difference between our mushroom and that of Dietz \etal
is that our foot supports BB orbits and theirs does not.
Therefore to eliminate this as a cause of difference,
in Fig.~\ref{fig:level}d we show the ergodic NNDS with BB modes
removed.
Here BB modes were identified as those with $f_j>0.7$ but small
integral on the base of the
foot, namely $\int_0^{1/2} (\mbf{n}\cdot\mbf{r}) |\nd{j}|^2 dq < 0.1$;
the BB subset comprises only 0.8\% of the total.
The difference between panels c) and d) is barely perceptible,
indicating that BB modes are not a significant contribution in our
setting.


Finally, in Fig.~\ref{fig:dos} we show the amplitude spectrum
$\tilde{\rho}(l) := \sum_{j=1}^n e^{ik_jl}$
of the density of states,
which highlights contributions from periodic orbits of length $l$.
Panel a) shows all levels, while b) and c) shows the
contribution only of
levels categorized as either regular or ergodic, according to the
above method.
The periodic peaks at the integers
in panel a) (and absent in b) are due to the BB mode in the foot.
As expected, b) contains only the regular clusters of peaks associated with
hat orbits which unfold to polygons in the disc.
Each cluster of peaks has an upper limit point
at multiples of $\pi R = 3\pi/2$ corresponding to whispering-gallery rays.
It is interesting that c) contains contributions not only from UPOs
but from all the peaks of b) too.

\section{Conclusion}
\label{sec:conc}

We have presented the first known high-lying eigenmode
calculations of Bunimovich's mushroom, which has unusually simple
divided phase space without KAM hierarchy.
Using a basis set adapted to the re-entrant corner,
the Method of Particular Solutions achieves very high
accuracy for low modes,
and the scaling method enables us to find high modes
orders of magnitude more efficiently than any other known numerical
approach, allowing 
the lowest $n=16061$ odd modes to be computed in reasonable time.
Since statistical estimation errors scale as $1/\sqrt{n}$,
we are therefore able to reach the 1\% level for many quantities.

Chaotic modes and Husimi functions
have been shown to be nonuniform and scarred
even at mode number $\approx 45000$,
evidence that the semiclassical limit is reached very slowly.
Using a separation into regular vs chaotic modes,
Percival's conjecture has been verified to within 2\%.
A new model for dynamical tunneling
(similar to that of B\"{a}cker \etal \cite{loeck07})
has been described, and shown
to predict the chaotic component of predominantly-regular modes
to within a factor of roughly an order of magnitude (over a range of $10^8$).
Its prediction (via Bessel asymptotics)
that the density of occurrence of
modes which are regular-chaotic superpositions dies
asymptotically like $E^{-1/3}$ agrees well with the first known
measurement of this density.

Our study of nearest-neighbor eigenvalue spacing
finds good agreement with the Berry-Robnik distribution,
and for the regular subset, good agreement with the Poisson distribution.
The ergodic subset shows statistically-significant deviations from
Wigner's GOE approximation, favoring small spacings.
However we find no evidence for the dip reported at $s=0.7$ 
by Dietz \etal \cite{dietz}; recall we study over 16 times their number of
modes.


This study is preliminary, and raises many interesting questions:
Can our model for dynamical tunneling be refined
to give agreement at the impressive level found in quantum maps \cite{loeck07}?
Does the ergodic level-spacing distribution eventually
tend to the GOE expectation?
Finally, can spectral manifestations of stickiness \cite{sticky1,sticky2}
be detected?


\acknowledgments
We thank Mason Porter, Eric Heller and Nick Trefethen
for stimulating discussions and helpful advice.
AHB is partially funded by NSF grant DMS-0507614.
TB is is supported by Engineering and Physical Sciences
Research Council grant EP/D079403/1.

\appendix
\section{Husimi transform} 
\label{app:hus}

We define the Husimi transform~\cite{stellar} of functions on $\mathbb{R}$,
for convenience reviewing the coherent state formalism
in dimensionless ($\hbar$-free) units.
Given a width parameter (phase space aspect-ratio) $\sigma>0$,
it is easy to show that the annihilation operator
\be
a := \frac{1}{\sqrt{2}}\left(\frac{q}{\sigma} + \sigma \partial_q\right)
\label{eq:ann}
\ee
has a kernel spanned by the $L^2$-normalized Gaussian
$\psi_0(q):=(\pi\sigma^2)^{-1/4} e^{-q^2/2\sigma^2}$.
We work in $L^2(\mathbb{R})$, in which the hermitian adjoint of $a$ is
$\cre = (q/\sigma - \sigma\partial_q)/\sqrt{2}$.
From the commutator $[a,\cre]=1$ it follows, $\forall z\in\mathbb{C}$, that
the coherent state
\be
\psi_z := e^{-|z|^2/2} e^{z\cre} \psi_0
\label{eq:coh}
\ee
is an eigenfunction of $a$ with eigenvalue $z$. The fact that it
is $L^2$-normalized requires the Hermite-Gauss normalization
$\| (\cre)^n \psi_0\|_2^2 = n!, \forall n\in\mathbb{N}$,
which can be proved by induction.
The Bargmann representation~\cite{bargmann1,bargmann2}
of a function $v:\mathbb{R}\to\mathbb{C}$
is then $\langle \psi_z, v\rangle$;
the Husimi representation is its squared magnitude
$H_{v,\sigma}(z):= |\langle \psi_z, v\rangle|^2$.
We need a more explicit form than (\ref{eq:coh}).
$\psi_z = e^{z\cre-z^*a}\psi_0$ follows
by the Baker-Campbell-Hausdorff formula $e^{A+B}=e^{-[A,B]/2}e^Ae^B$
for $[[A,B],A] = [[A,B],B] = 0$.
Applying this formula again and writing
$z := (q_0/\sigma + i
\sigma k_0)/\sqrt{2}$ where $q_0,k_0\in\mathbb{R}$ 
gives
\be
\psi_z(q) = e^{ik_0q_0/2} e^{ik_0q} \psi_0(q-q_0).
\label{eq:exp}
\ee
This shows that the coherent state is localized in position (around $q_0$) and
wavenumber (around $k_0$),
thus the Husimi is a microlocal (phase space) represention,
\be
H_{v,\sigma}(q_0,k_0) := \left| \int_{-\infty}^\infty
v(q) e^{ik_0q} \psi_0(q-q_0) dq \right|^2.
\label{eq:hus}
\ee
This also known as the Gabor transform or spectrogram (windowed Fourier
transform), and it can be proven
equal to the 
Wigner transform convolved by the smoothing function $\psi_0^2$.
Given a normal-derivative function $\nd{j}$ we periodize it
in order to apply the above. We also scale the wavenumber by $k_j$,
thus the Birkhoff momentum coordinate is $p=k_0/k_j$.

\section{Integrable phase-space fraction}
\label{app:uint}
 
The total phase space (restricting to the unit-speed momentum shell)
has volume $V_{\tbox{tot}} = \vol(\rO\times S^1) = 2\pi\vol\rO =
2\pi(ab/2 + \pi R^2/4)$.
Define the function
$\alpha\ofr:= 2\pi - 4\sin^{-1}(b/2d\ofr)$, where $d\ofr$ is the distance
from $\mbf{r}$ to the center point $q_c$.
When $\mbf{r}$ is in the hat and $d\ofr\in[b/2,R]$,
$\alpha\ofr$ gives the measure of the set of angles in $S^1$
for which orbits launched from $\mbf{r}$ are integrable
(\ie never leave the annulus $d\ofr\in[b/2,R]$).
The regular phase space volume is found by integrating $\alpha\ofr$
over the quarter-annulus using polar coordinates
$(\rho,\phi)$:
\bea
V_{\tbox{reg}} &= &\int_0^{\pi/2} d\phi \int_{b/2}^R \alpha(\rho)\rho d\rho
\nonumber \\
&=&
\frac{\pi^2}{2}\left(R^2-\frac{b^2}{4}\right)
-2\pi\int_{b/2}^R \rho \sin^{-1} \frac{b}{2\rho} \, d\rho
\nonumber \\
&=&
\pi R^2\left(\cos^{-1}\frac{b}{2R} \;- \;\frac{b}{2R} p_0\right)~.
\nonumber
\eea
The same result is given without calculus using the space
of oriented lines in a full annulus, that is, $4V_\tbox{reg} = 2\pi$
times the area of the segments $\{(x,y): x^2+y^2<R^2, |y|>b/2\}$.
For our parameters we get $\mu_\tbox{reg} := V_{\tbox{reg}}/V_{\tbox{tot}}
= 0.4549\cdots$

\bibliographystyle{apsrev} 
\bibliography{alex}

\begin{figure}[p]
\bc
\includegraphics[width=\textwidth]{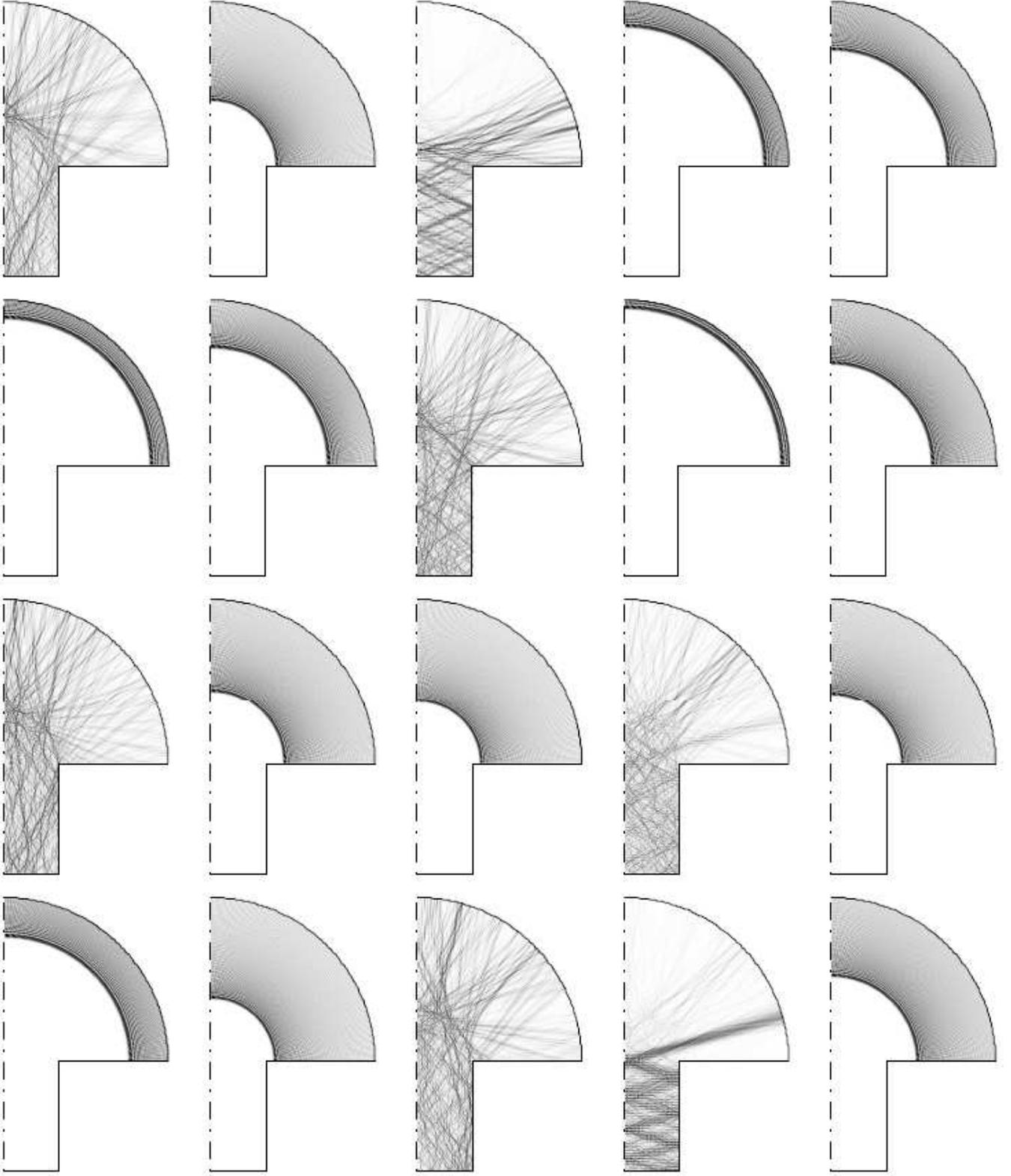}
\ec
\caption{
20 high-eigenvalue consecutive modes,
covering the range $k_j \in [499.800, 499.869]$,
with mode number $j\approx 45000$.
Mode number increases horizontally from the top
left. $|\phi_j|^2$ is shown with zero white and larger values darker.
\label{fig:many}
}
\end{figure}

\begin{figure}[p]
\bc
\includegraphics[width=\textwidth]{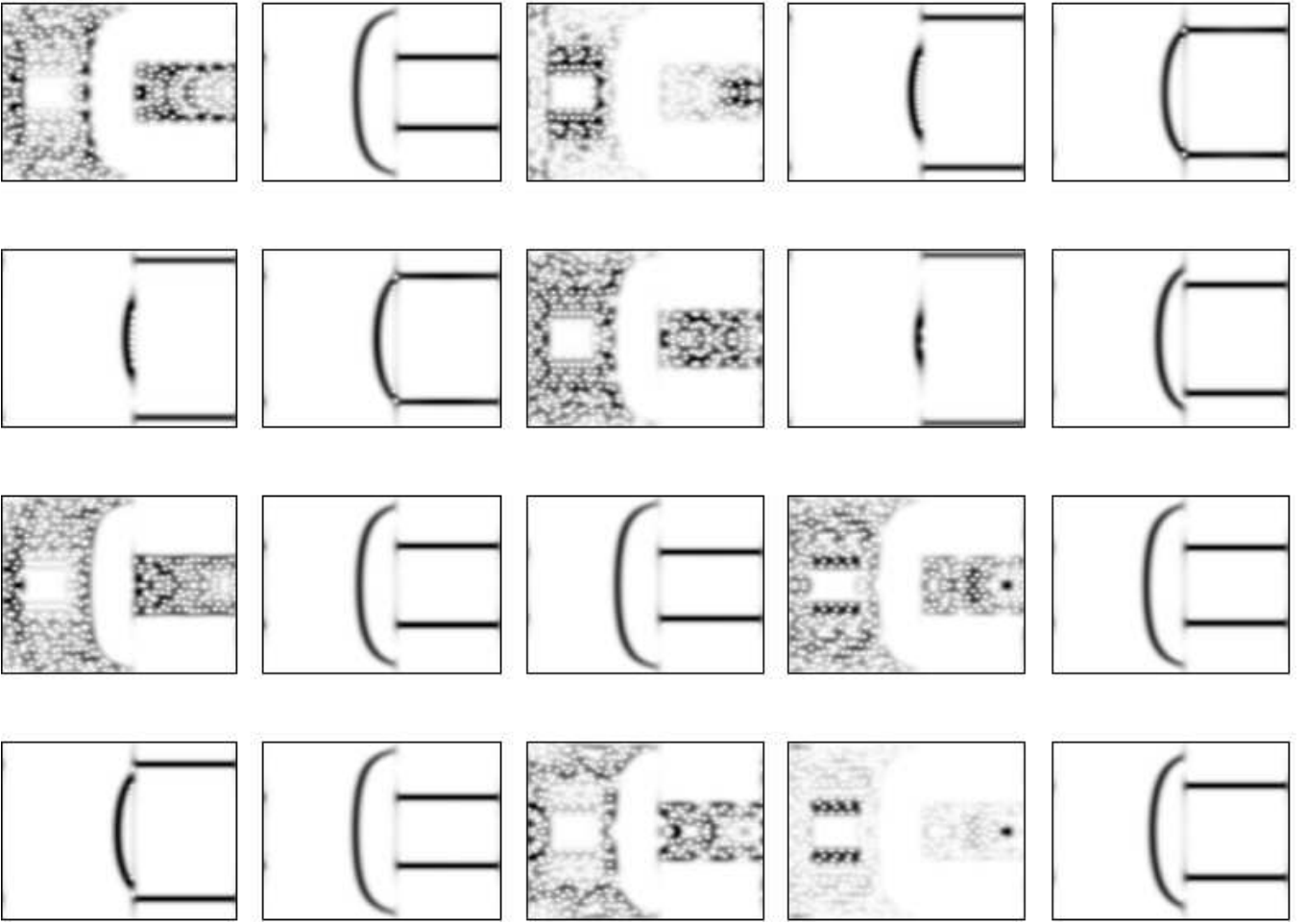}
\ec
\caption{
Husimi distributions $H_{\nd{j},\sigma}(q,p)$ of the
20 high-eigenvalue modes shown in Fig.~\ref{fig:many}, and in the same order.
The $q$ and $p$ axes are as in Fig.~\ref{fig:hus}.
\label{fig:manyhus}
}
\end{figure}

\end{document}